\documentclass[11pt]{article}

\usepackage{sectsty}
\usepackage{graphicx}

\usepackage{mathtools}
\DeclarePairedDelimiter{\ev}{\langle}{\rangle}

\title{One rule does not fit all: deviations from universality in human mobility modeling}
\author{Ludovico Napoli$^1$, M\'arton Karsai$^{1,2}$, Esteban Moro$^{3}$\footnote{Corresponding author: \texttt{e.moroegido@northeastern.edu}}}
\date{}

\begin{document}
\maketitle	


\begin{center}
$^1$ Department of Network and Data Science, Central European University, Quellenstraße 51, 1100 Vienna (Austria) \\
$^2$ National Laboratory for Health Security, Alfréd Rényi Institute of Mathematics, Re\'altanoda utca, Budapest 1053, (Hungary) \\
$^3$ Network Science Institute, Northeastern University, 177 Huntington Ave, Boston, MA 02115 (USA) \\
\end{center}

\begin{center}
    \textbf{Abstract} \\

\end{center}
The accurate modeling of individual movement in cities has significant implications for policy decisions across various sectors. Existing research emphasizes the universality of human mobility, positing that simple models can capture population-level movements. However, population-level accuracy does not guarantee consistent performance across all individuals. By overlooking individual differences, universality laws may accurately describe certain groups while less precisely representing others, resulting in aggregate accuracy from a balance of discrepancies. Using large-scale mobility data, we assess individual-level accuracy of a universal model, the Exploration and Preferential Return (EPR), by examining deviations from expected behavior in two scaling laws — one related to exploration and the other to return patterns. Our findings reveal that, while the model can describe population-wide movement patterns, it displays widespread deviations linked to individuals’ behavioral traits, socioeconomic status, and lifestyles, contradicting model assumptions like non-bursty exploration and preferential return. Specifically, individuals poorly represented by the EPR model tend to visit routine locations in sequences, exploring rarely but in a bursty manner when they do. Among socioeconomic factors, income most strongly correlates with significant deviations. Consequently, spatial inhomogeneity emerges in model accuracy, with lower performance concentrated in urbanized, densely populated areas, underscoring policy implications. Our results show that emphasizing population-wide models can propagate socioeconomic inequalities by poorly representing vulnerable population sectors.
\pagebreak

\section*{Introduction}

As the global urban population continues to rise, cities play an increasingly pivotal role in societal development and daily life. This demographic shift amplifies the need for a deep understanding of human mobility within urban environments. Effective modeling of city movements is crucial not only for urban planning \cite{kitamura2000micro} and infrastructure development \cite{barthelemy2016structure}, but also for enhancing public services \cite{wang2020review}, alleviating traffic congestion \cite{nagel1995emergent, ccolak2016understanding}, fostering social integration \cite{moro2021mobility}, and controlling the spread of epidemics \cite{balcan2009multiscale, tizzoni2014use}.

Human mobility models \cite{barbosa2018human} often seek to identify and replicate universal laws governing movement patterns. These include the distribution of travel distances \cite{brockmann2006scaling, alessandretti2020scales}, the scaling of mobility flows with distance and population size \cite{zipf1946p, simini2012universal}, and the spatiotemporal dynamics of aggregate movements \cite{schlapfer2021universal}. While these models have made significant strides in replicating observed behavioral patterns, their universal applicability remains uncertain across diverse population groups. It is still unclear whether these models represent all individuals equally or favor certain demographic characteristics, potentially marginalizing others. Addressing this issue is essential to ensure that mobility models equitably support inclusive urban development.

Over the years, various approaches have been proposed to model human mobility at both population and individual levels. At the macro level, mobility models often rely on the gravity \cite{zipf1946p, erlander1990gravity, simini2021deep} and radiation \cite{simini2012universal} laws to predict movements between meta-populations. At the individual level, random walk-based models \cite{brockmann2006scaling, toole2015coupling} have been employed, assuming that humans behave like randomly moving particles \cite{montroll1965random}. The Exploration and Preferential Return (EPR) model \cite{song2010modelling} stands out among individual mobility models due to its simplicity and ability to replicate key scaling laws observed in human mobility. By assuming lower stochasticity in human trajectories, this model increases predictability \cite{gonzalez2008understanding, song2010limits} by focusing on two primary mechanisms: people explore new places less frequently over time and revisit highly visited locations more often. Due to its robustness, the EPR model has inspired extensions incorporating factors such as physical constraints \cite{pappalardo2016human}, recency effects \cite{barbosa2015effect}, ranking mechanisms \cite{jiang2016timegeo}, place relevance \cite{pappalardo2015returners}, routine behaviors \cite{pappalardo2018data}, and social preferences \cite{moro2021mobility}. However, the model assumes that all individuals follow the same mobility mechanisms, overlooking variations in mobility patterns across different socioeconomic groups and lifestyles. For instance, socioeconomic status influences daily travel distance \cite{xu2018human, leo2016socioeconomic, lotero2016rich}, travel frequency \cite{barbosa2021uncovering}, the socioeconomic profile of visited locations \cite{hilman2022socioeconomic, dong2020segregated, moro2021mobility}, and the ability to adjust mobility during emergencies \cite{gozzi2021estimating, pullano2020evaluating, duenas2021changes, napoli2023socioeconomic}. Additionally, people visit places according to identifiable activity behaviors and lifestyles that span across sociodemographic groups \cite{yang2023identifying}. These factors likely affect the balance between exploration and recurrent visits, which are not considered by the EPR model, potentially leading to varying levels of accuracy across demographic and lifestyle groups.

In this paper, we aim to evaluate the performance of the EPR model at the individual level. Our focus is on mobility scaling laws related to exploration dynamics and visitation frequencies, assessing whether the model captures the mobility of individuals with different characteristics equally well. Our analysis reveals a broad range of deviations from the model's expectations, arising from violations of its microscopic assumptions, particularly non-bursty exploration and preferential return. We investigate the types of places visited when these assumptions are violated and the sociodemographic and lifestyle characteristics of individuals who deviate most from the model. Finally, we discuss the key implications of our findings, highlighting the consequences of unequal model representation across different population groups, despite the model’s perceived universality.

\section*{Results}

To address our scientific challenge, we analyze the micro-scale movements of 1.5 million anonymized users in 11 core-based statistical areas \cite{cbsa} (CBSAs) in the US, recorded between October 2016 and March 2017. These data are extracted from a device-level dataset of high-resolution privacy-peserving mobile location pings for 4.5 million devices (for details on the raw data and cleaning process, see Methods). For each user, we have a list of fine-grained spatially and temporally localized visits, each associated with a Foursquare venue and its category classification \cite{foursquare}. In the following analysis, we focus on a subsample of 51,648 users from the Boston-Cambridge-Newton CBSA (referred to as Boston for simplicity). The analysis of other CBSAs can be found in the Supplementary Information (SI).

\begin{figure*}[ht!] 
\includegraphics[width=\linewidth]{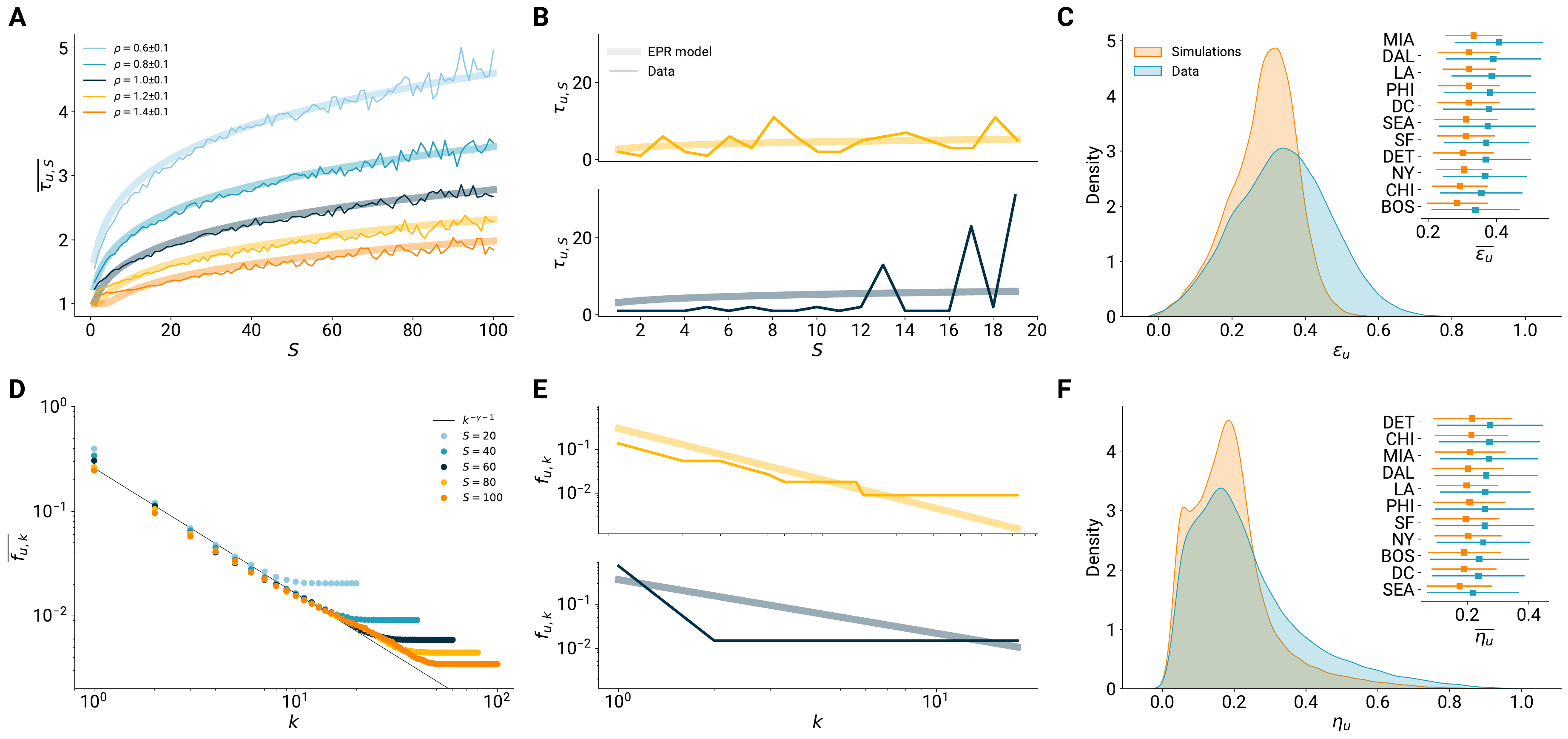} 
\caption{\textbf{Individual deviations beyond population-wide accuracy.} A) The average exploration inter-event time as a function of the number of distinct places $S$ for users with different values of $\rho_u$ (thin lines) and the corresponding predictions by the EPR model (thick lines). B) Individual-level observations of exploration inter-event time as a function of $S$ (thin lines) and the predictions of the EPR model (thick lines) for well-fitting (upper panel) and deviating (lower panel) behavior compared to the model. C) Distribution of individual deviations $\epsilon_u$ from the data (light blue) and from stochastic simulations of the EPR model (orange). The inset shows the mean and standard deviations of the same distributions across all CBSAs. D) Average visitation frequency of ranked locations according to the EPR model (thick black line) and for users with different values of $S$. E) Same as B) but for visitation frequency. F) Same as E) but for $\eta_u$.} 
\label{fig1} 
\end{figure*}

\subsection*{Individual level deviations}

In the framework of the Exploration and Preferential Return (EPR) model, users move between locations. At each step, a user $u$ can either return to a previously visited place (return step) or explore a new place (exploration step). The individual tendency of a user $u$ to explore, given that they have already visited $S$ distinct locations, is embedded in the exploration probability $P_u(S) = \rho_u S^{-\gamma}$. This probability is controlled by the individual parameter $\rho_u$, which is derived from the total number of distinct places visited by $u$ compared to the total number of visits (see Methods for details). The parameter $\gamma = 0.22$ is a global constant, as determined in \cite{moro2021mobility}.

The EPR model accurately reproduces two key aspects of human mobility: exploration dynamics and visitation frequency, at least on average. For exploration dynamics, we focus on the inter-event time $\tau_{u,S}$, which we define as the number of steps after which a user $u$ visits a new location, given that they have already visited $S$ distinct places. According to the model (see Methods for details), the expected inter-event time $\ev{\tau_{u,S}}$ is given by $\ev{\tau_{u,S}} = P_u^{-1}(S) = S^{\gamma}/\rho_u$. Meanwhile, the visitation frequency to the $k$th most visited location by a user $u$ follows Zipf's law, where $\ev{f_{u,k}} \sim k^{-\gamma}$.

If we average $\tau_{u,S}$ or $f_{u,k}$ among users in our data, as seen in panels Fig.\ref{fig1} A and D, the EPR model accurately reproduces both the exploration inter-event time and the visitation frequency. However, correctly representing the average behavior does not necessarily imply an equally precise description of all individuals in the observed population. Since both properties can be measured for each user, we can assess the precision of the model at the individual level. 

For demonstration, Fig.\ref{fig1} B and E (upper panels) show an example of a user whose $\tau_{u,S}$ and $f_{u,k}$ adhere well to the scaling laws. In contrast, Fig.\ref{fig1} B and E (lower panels) illustrate the same metrics for a user whose behavior deviates from the predicted scaling. To quantify this effect, we introduce two individual metrics that capture deviations from the expected scaling of the inter-event time and the visitation frequency. For the exploration inter-event times, we define $\epsilon_u$ as the symmetric mean absolute percentage error (SMAPE) between $\tau_{u,S}$ and $\ev{\tau_{u,S}}$ for a user $u$:
$$
\epsilon_u = \frac{1}{S_u} \sum_{S=1}^{S_u} \frac{|\tau_{u,S} - \ev{\tau_{u,S}} |}{|\tau_{u,S}| + |\ev{\tau_{u,S}}|},
$$
where $S_u$ is the total number of distinct places visited by the user $u$. For the visitation frequency, $\eta_u$ is defined as the Kullback-Leibler deviation between $f_{u,k}$ and $\ev{f_{u,k}}$:
$$
\eta_u = \sum_{k=1}^{K_u} f_{u,k} \log \frac{f_{u,k}}{\ev{f_{u,k}}},
$$
where $K_u$ is the rank of the least visited location (excluding locations visited only once, to remove the tail effect visible in panel B of Fig.\ref{fig1}). Both of these metrics can be computed from real data as well as from simulated mobility trajectories generated by the EPR model, to assess if the observed deviations are comparable to the ones of EPR-based stochastic trajectories.

The distributions of $\epsilon_u$ and $\eta_u$ for real and simulated users in the Boston area are shown in Fig.\ref{fig1} C and F, respectively. To facilitate comparison between the two variables (which is necessary for subsequent analysis), we normalize them to a range between 0 and 1. These distributions reveal that, while the EPR model reproduces the exploration dynamics and visitation frequency well on average, it does not equally represent the behavior of every user. Notably, the observed deviations are not solely due to the stochastic nature of the model. The orange distributions in Fig.\ref{fig1} C and F show the equivalent results from stochastic EPR model simulations for the same set of users (see the SI for details). Although stochasticity partially accounts for the observed variance (particularly for $\eta_u$), significant differences between the data and simulation-driven distributions remain, particularly for the largest deviations, which cannot be explained by the model's stochastic variation alone. This observation holds not only for Boston but for every CBSA considered, as demonstrated in the subpanels of Fig.\ref{fig1} C and F, which display the mean and standard deviations of the real and simulated distributions of $\epsilon_u$ and $\eta_u$ across all CBSAs.

Interestingly, despite measuring different mobility properties, the two deviation metrics show a strong correlation (Pearson correlation of 0.75). This indicates that users who are poorly represented by the model in terms of exploration dynamics are also likely to be poorly represented in terms of visitation frequency (see the SI for further details).

\begin{figure*}[ht!]
  \includegraphics[width=\linewidth]{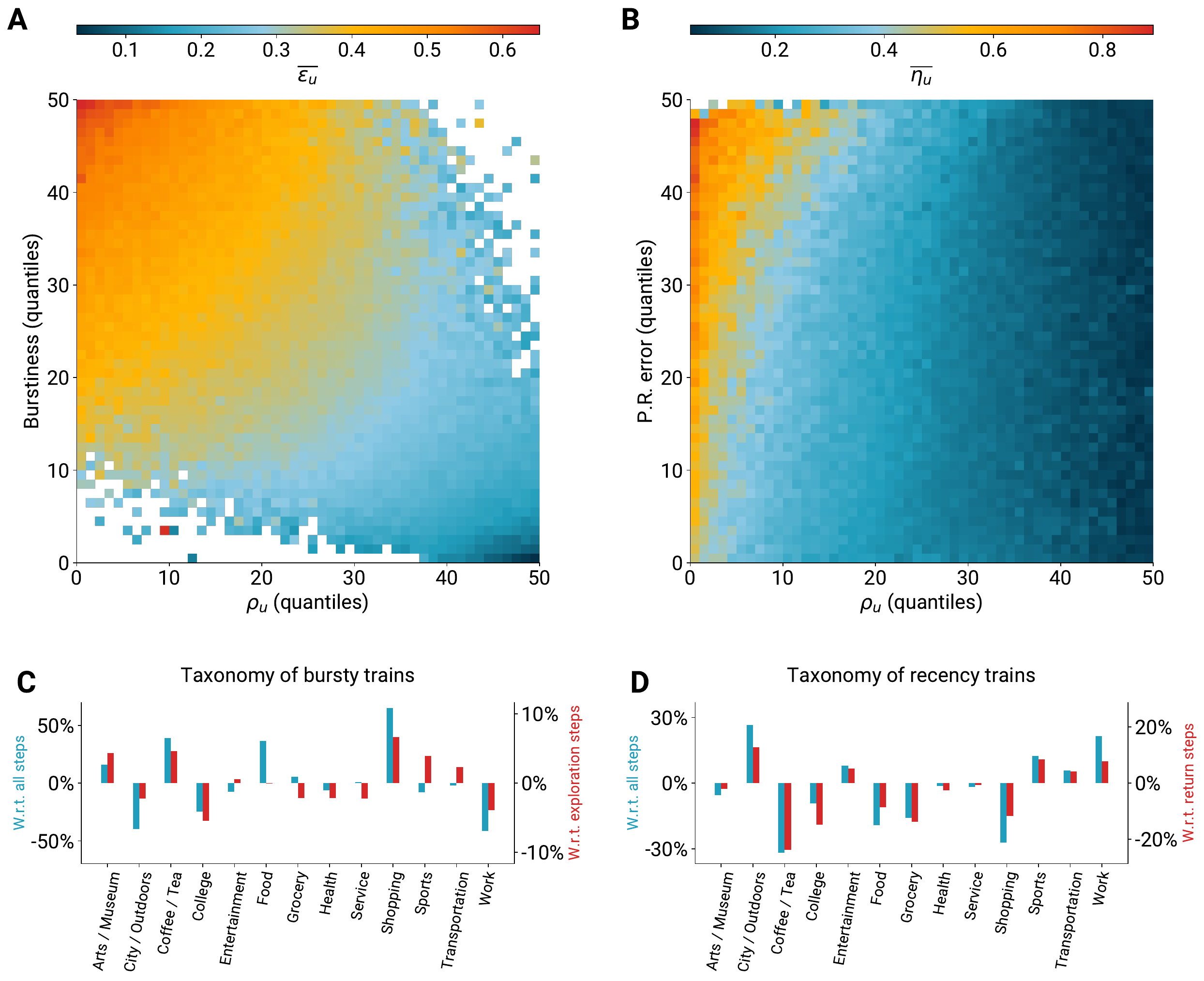}
\caption{\textbf{Deviations are related to violations of the EPR model's microscopic mechanisms.} A) Average values of the deviation $\epsilon_u$ (color-coded as in the color bar) for users grouped in quantiles of $\rho_u$ and exploration burstiness. Controlling for stochasticity with $\rho_u$, $\epsilon_u$ increases with burstiness. B) Average values of the deviation $\eta_u$ (color-coded as in the color bar) for users grouped in quantiles of $\rho_u$ and preferential return error (P.R. error). Controlling for stochasticity with $\rho_u$, $\eta_u$ increases with P.R. error. C) Characterization of bursty trains, i.e., sequences of consecutive exploration steps, in terms of relative visits to defined categories, compared to all visits (blue bars, left y-axis) and to exploration steps only (red bars, right y-axis). D) Characterization of recency trains, i.e., sequences of consecutive visits to the same place, in terms of relative visits to defined categories, compared to all visits (blue bars, left y-axis) and to return steps only (red bars, right y-axis).}
\label{fig2}
\end{figure*}

\subsection*{Microscopic mechanisms}

Having seen the wide range of deviation metrics that cannot be fully explained by stochasticity, our aim is to understand the microscopic characteristics associated with these deviations. More specifically, we seek to identify and investigate the mobility aspects that the EPR model, due to its assumptions, does not account for, and which may lead to poor representativity.

First, we expect the deviations to be related to the individual exploration tendency, encoded in the parameter $\rho_u$. Indeed, $\rho_u$ controls the stochastic part of both $\epsilon_u$ and $\eta_u$, which influences the distribution of these quantities as generated from simulations, as shown in Fig.\ref{fig1} C and F (for details, see SI). However, despite its clear relationship with $\epsilon_u$ and $\eta_u$, the exploration tendency is already integrated into the EPR model, accounting for a certain level of uncertainty due to the model's stochastic nature. Here, we focus on the characteristics or mechanisms that the model fails to capture. By using the exploration tendency as a control variable, we can rule out the stochastic component of the deviations. This simplifies our primary question to: \emph{given a sample of users with the same exploration tendency (i.e., with the same associated stochastic uncertainty), what individual microscopic characteristic makes some users more accurately represented than others, i.e., with smaller observed deviations from the model?}

Regarding the inter-event times between consecutive explorations, the EPR model assumes that the probability of exploration decreases uniformly as the number of distinct places grows, according to the relation $P_u(S) = \rho_u S^{-\gamma}$. This assumption about the dynamics neglects burstiness, i.e., the interplay between long periods of low activity and short periods of high activity, a widely observed phenomenon in human dynamics \cite{karsai2018bursty}. We hypothesize that users who exhibit more bursty exploration patterns, i.e., those who alternate between short periods of intense exploration and long periods of inactivity, will have a higher $\epsilon_u$. These users exhibit a stronger tendency towards a microscopic feature that is not captured by the EPR model and is closely related to the inter-event time $\tau_{u,S}$. 

To test this hypothesis, we measure the burstiness parameter as defined in \cite{kim2016measuring} for each user (see Methods for details) and analyze its relation with $\epsilon_u$, while controlling for stochastic uncertainty via $\rho_u$. The results are shown in Fig.\ref{fig2} A, where we group users into quantiles of $\rho_u$ and burstiness and compute the average $\epsilon_u$ within each group. This parameter space reveals that, beyond the $\rho_u$ dependency, $\epsilon_u$ is positively associated with burstiness. In fact, when considering a sample of users with the same exploration tendency, the more bursty users tend to be less well-represented by the EPR model in terms of their exploration dynamics.

The microscopic mechanism leading to the distribution $\ev{f_k} \sim k^{-\gamma}$ of visitation frequency in the EPR model is the preferential return criterion. This is a widely observed and modeled phenomenon in human systems \cite{barabasi1999emergence}, though it may not equally describe the behavior of all individuals. According to this criterion, the probability $\Pi_i$ of visiting location $i$ is equal to the fraction of visits $\phi_i$ the user has made to that location up to the point of observation. Our hypothesis is that the less a user follows this criterion, the less their visitation frequency distribution will align with the $\ev{f_k} \sim k^{-\gamma}$ scaling. 

To test this hypothesis, we measure the error associated with the preferential return criterion (P.R. error) as the SMAPE between $\Pi_i$ and $\phi_i$ for each user (for details, see Methods). Similar to the case of burstiness, we group users into quantiles of $\rho_u$ and P.R. error, and compute the average $\eta_u$ deviation within each group. As shown in Fig.\ref{fig2} B, the deviation $\eta_u$ is positively associated with the P.R. error, in addition to its expected dependence on $\rho_u$, especially for the lowest values of $\rho_u$. Specifically, for users with similarly low exploration tendencies, the more they deviate from the preferential return mechanism, the less their visitation frequency distribution aligns with the EPR model's expected outcome of $\ev{f_k} \sim k^{-\gamma}$. Conversely, for highly explorative users, adherence to the preferential return mechanism does not seem to strongly influence the accuracy of the visitation frequency predicted by the EPR model.

To determine whether these observations are generalizable to other locations, we repeated our analysis for all CBSAs. As it turns out (see SI for details), the roles of burstiness and preferential return in determining the deviations $\epsilon_u$ and $\eta_u$ are robust and consistent across all CBSAs analyzed.

\subsection*{Characterization of assumptions' violations}

After assessing the direct and individual-level relationship between deviations from scaling laws and violations of the EPR model's assumptions, we aim to characterize these violations from a behavioral perspective. In other words, we want to investigate what types of places people visit when they deviate from the smooth exploration dynamics of the EPR model, instead engaging in bursty exploration periods. Similarly, when people violate the preferential return criterion, where do they tend to return?

We can address these questions because, as previously mentioned, every step in our dataset is associated with a Foursquare category and one of 15 macro taxonomy groups (see Methods for details), such as \textit{Coffee/Tea, Grocery}, and \textit{Shopping}. To characterize exploration burstiness, we analyze the places visited during bursty exploration trains, i.e., sequences of consecutive exploration steps. Specifically, we compare the frequency of visits to a given category with its overall frequency of visits, as well as its frequency of visits during exploration steps only. Both comparisons are important: the first tells us how much more or less a category is visited during bursty trains compared to any other step, while the second restricts the comparison to exploration steps only, as certain categories may generally be visited more frequently during exploration but not necessarily during bursty periods. 

As shown in Fig.\ref{fig2} C, bursty trains are indeed characterized by specific categories. In particular, amusement places such as museums and art galleries, coffee shops, and shopping locations are visited significantly more often during bursty exploration trains. Conversely, routine locations such as transportation hubs and workplaces are rarely explored during bursty trains. Additionally, the macro category \textit{City / Outdoor}, which comprises parks, neighborhoods, and residential places, is typically not explored in a bursty manner. The other categories show either small or diverging differences with respect to the comparisons considered.

Regarding violations of preferential return, recency has been identified and observed as one of the phenomena that characterize human mobility and is not captured by the EPR model \cite{barbosa2015effect}. Recency describes a memory effect, indicating that humans tend to return not only to highly visited locations but more frequently to those visited recently. In contrast, the preferential return criterion does not account for this effect, assuming that all locations can be revisited based solely on their past frequency of visits, without considering the temporal order of those visits. To capture this effect while drawing a parallel with bursty trains, we introduce the concept of recency trains, i.e., sequences of consecutive return visits to the same place. We characterize these recency trains using the same methodology applied to bursty exploration trains. Specifically, we compare the frequency of visits to a given category with both its overall frequency and, in this case, its frequency during return steps only.

The results are shown in Fig.\ref{fig2} D, where we observe a somewhat mirrored pattern compared to bursty trains. When people repeatedly return to the same place, they tend to do so at routine and habitual locations, such as residential areas, workplaces, as well as transportation hubs and sports venues. In contrast, amusement places like coffee shops, restaurants, and shopping malls are not typically revisited continuously. Interestingly, entertainment places also appear in recency trains, though not significantly. It is also surprising that routine categories such as \textit{College} and \textit{Groceries} are notably absent from recency trains.

The results of this section are consistent across all CBSAs, as detailed in the SI.

\begin{figure*}[ht!]
  \includegraphics[width=\linewidth]{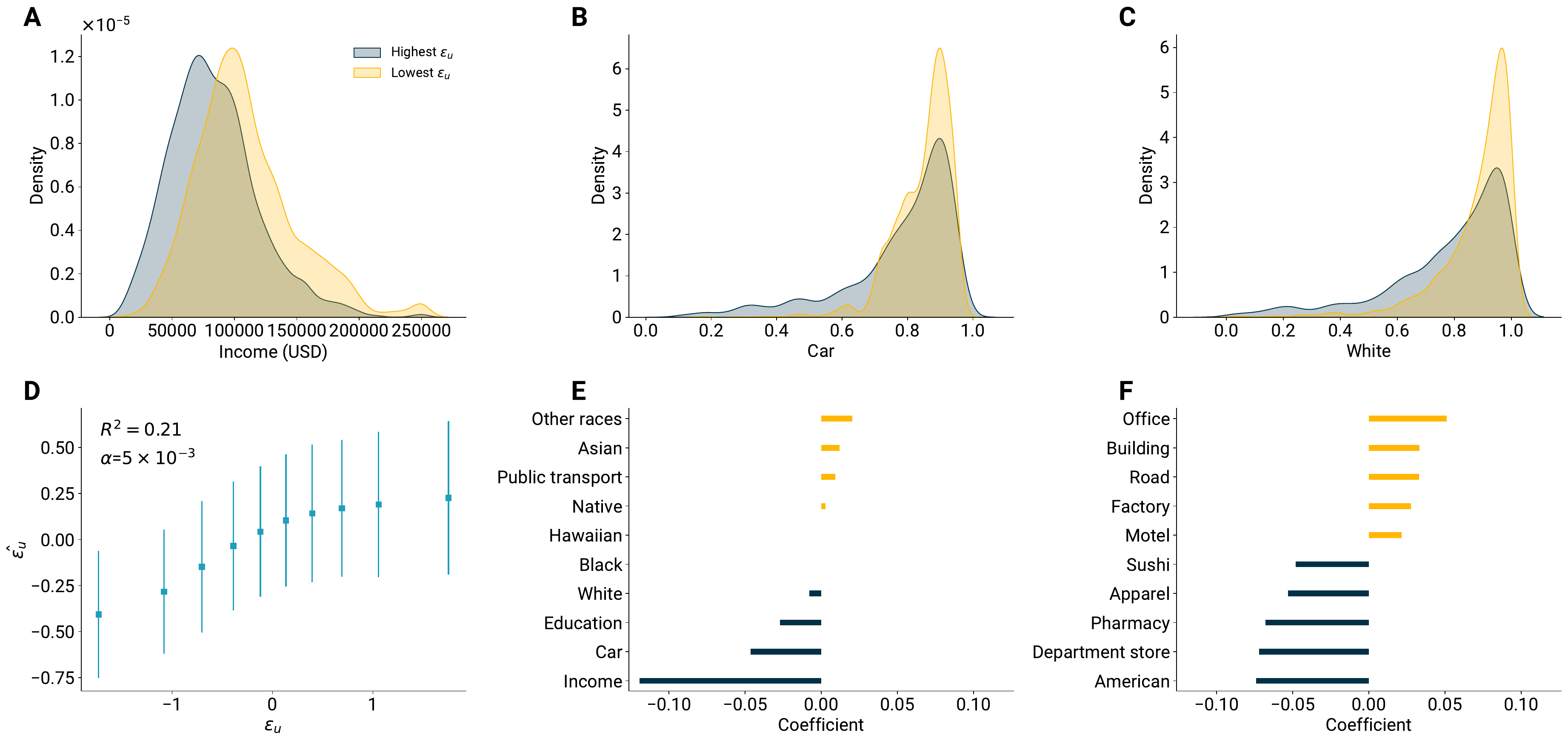}
\caption{\textbf{Deviations are biased towards sociodemographics and lifestyles.} A) The income distribution of users in the highest (blue) and lowest (yellow) 10\% quantiles of the deviation $\epsilon_u$. B) Same as A) for car usage. C) Same as A) for the probability of being white. D) Results of the LASSO regression: true $\epsilon_u$ vs predicted $\hat{\epsilon_u}$ based on sociodemographic and lifestyle features. $R^2$ is the coefficient of determination, and $\alpha$ is the regularization parameter. E) Coefficients of the socioeconomic features from the LASSO regression. F) Coefficients of the life habit features (showing only the highest and lowest 5) from the LASSO regression.}
\label{fig3}
\end{figure*}

\subsection*{Vulnerable groups}

Finally, having identified the microscopic mechanisms whose violations are associated with deviations from the model, and how these violations are characterized from a behavioral perspective, we aim to characterize who is more prone to these patterns and, hence, at risk of not being modeled correctly. To answer this question, we test whether the deviations are related to any individual traits concerning the sociodemographics and life habits of people. Our goal is to explore whether deviations are uniformly distributed across the population or if there are specific groups that show significantly larger deviations, meaning they are less well-represented by the EPR model. Based on their inferred home census tract, we assign users with sociodemographic indicators (related to wealth, education, race, and means of transportation, see Methods for details) and lifestyle indicators (the fraction of times they visited places of a given category, see Methods for details).

Given a sociodemographic indicator, we compute its distribution for people in the highest and lowest 10\% quantiles of the $\epsilon_u$ and $\eta_u$ distributions. These are users who are respectively described as the worst and best by the EPR model. In Fig.\ref{fig3}, we show these distributions for the sociodemographic variables of income (panel A), car usage (B), and the probability of being white (C) for the two extreme quantiles of $\epsilon_u$. The income distribution of the worst-described users in the highest 10\% of $\epsilon_u$ is shifted to lower values compared to the best-described users in the lowest 10\%. This implies a possible correlation between income and the $\epsilon$ deviation, suggesting that the EPR model predicts the mobility of people with lower incomes less accurately. Meanwhile, the same distributions computed for white people and car owners are less concentrated but skew towards lower values for those in the highest 10\% of $\epsilon$ errors. Additionally, similar observations appear when investigating the $\eta_u$ deviations (for details, see SI). These findings suggest that users poorly represented by the EPR model, in terms of both exploration dynamics and visitation frequency, are more likely to be poorer, less likely to own a car, and less likely to be of white racial origin than those who are well-represented by the model.

To establish a more robust observation of these sociodemographic biases, not only for extreme quantiles but for all users, we perform a LASSO regression with 10-fold cross-validation. We use the deviation metrics ($\epsilon_u$ or $\eta_u$) as dependent variables and all sociodemographic and life habit features as predictors (see Methods for details). We control for $S_u$ and $K_u$ (respectively for $\epsilon_u$ and $\eta_u$) to avoid measuring an effect due to varying sample sizes used for the computation of the dependent variables. Interestingly, we achieve $R^2=0.21$ correlation between the observed $\epsilon_u$ and predicted $\hat{\epsilon}_u$ ($R^2=0.22$ for $\eta_u$, see SI), indicating that the deviations are not randomly distributed. On the contrary, they can be partially predicted by the sociodemographic and life habit features of users, as seen in Fig.\ref{fig3} D and the SI. Similar scores are achieved for all 11 CBSAs (see SI for details).

The different features ranked by their regression coefficients (in Fig.\ref{fig3} E) identify income as the most relevant sociodemographic characteristic in determining the $\epsilon$ error (and also $\eta$, see SI). Indeed, this significantly negative coefficient of income in both regressions verifies our earlier observation that users who are poorly represented by the model are more likely to be poorer. Moreover, this observation is robust across all CBSAs analyzed (for details, see SI), with income consistently showing the largest negative coefficient. Exceptions appear for $\eta_u$ in New York and Los Angeles, where income has the second most negative coefficient (details in the SI). On the other hand, the coefficient for car use behaves differently for the two deviation metrics, being negatively associated with $\epsilon_u$ and positively with $\eta_u$, not only in Boston but in most CBSAs (details in SI). Finally, contrary to our earlier conjecture, ethnicity does not strongly influence model performance in Boston, but it is significant in some CBSAs (details in SI). In particular, the probability of being African American is negatively associated with deviations in many CBSAs, indicating that Black individuals are often more likely to be well-represented by the model.

Similar to the sociodemographic variables, model deviations are also strongly biased concerning certain life habits, as shown in Fig.\ref{fig3} F, where we present five features with the highest and lowest coefficients (for more results, see SI). Places like offices, factories, and buildings are primarily positively associated with larger deviations. In contrast, negatively correlated categories include Sushi, American, and Fast Food restaurants, shopping malls, and amusement places like Movie Theaters. Similar patterns are observed across all CBSAs (see SI for details). Overall, users with more work-driven daily routines, who spend more time in offices and factories, are more likely to be poorly represented by the model, while users who spend more time in restaurants and amusement places are less likely to experience deviations.

\begin{figure*}[ht!]
  \includegraphics[width=\linewidth]{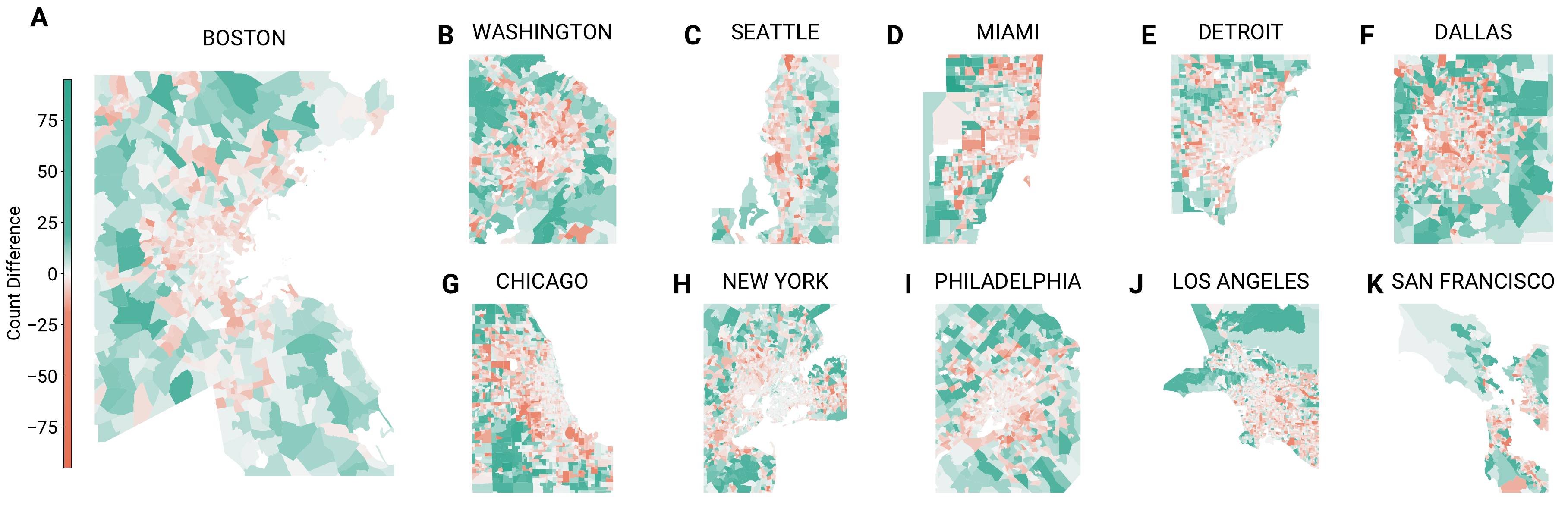}
\caption{\textbf{Urban-rural divide.} A-K) U.S. census tracts division of the 11 CBSAs under consideration, where each tract is colored according to the difference count between the number of users in the highest and lowest 10\% of $\epsilon_u + \eta_u$ with estimated home location in the tract.}
\label{fig4}
\end{figure*}

\section*{Discussion}

Our study highlights the limitations of evaluating human mobility models based solely on population-wide behavior and the risks of assuming universality across diverse populations. By analyzing large-scale mobility data, we compared individual exploration dynamics and visitation frequency with the predictions of the EPR model, revealing significant variations in adherence that cannot be fully explained by stochasticity. These findings underscore the potential pitfalls of assuming that a single framework applies equally to all individuals. In reality, urban mobility is highly heterogeneous, influenced not only by personal preferences but also by external constraints. While unified models can offer valuable insights at the population level, it is essential to recognize that their accuracy will not be uniform across all individuals, and may be particularly poor for some.

Our results demonstrate that individuals who deviate from the EPR model share distinct behavioral patterns. They tend to visit habitual locations in consecutive sequences and rarely engage in exploration. When they do explore, it happens in short bursts, predominantly at non-routine and amusement places. These behaviors are not randomly distributed across the population but are strongly linked to socioeconomic factors, with lower-income individuals being particularly poorly represented by the model. This bias compounds the substantial inequities already present in human mobility studies, where vulnerable groups are typically under-represented in digital data \cite{omodei2022complex}. Such misrepresentation has serious consequences for policy-making and modeling \cite{schlosser2020covid, sekara2023machine}, potentially exacerbating existing inequalities \cite{verma2019weapons}. Furthermore, during the COVID-19 pandemic, the mobility of low-income individuals was less adaptable due to limited capacity to stay at home, leading to higher exposure to infection and mortality \cite{mena2021socioeconomic, pullano2020evaluating, gozzi2021estimating, duenas2021changes}. The unequal representation in mobility models, as revealed by our study, poses additional risks for these vulnerable populations. Addressing these factors is critical for achieving equitable human mobility modeling.

A notable implication of our findings is the unequal spatial accuracy of the EPR model. As shown in Fig. \ref{fig4}, users who are better represented by the model tend to live in less densely populated suburban and rural areas, whereas those with the highest deviations are concentrated in densely populated urban environments (see SI for details). This spatial pattern reflects residential segregation by income, with wealthier individuals typically residing outside of city centers. Paradoxically, the EPR model, designed to simulate urban mobility, performs better in rural settings where movement patterns are more predictable.

From a policy-making perspective, this poses significant challenges. Urban environments, characterized by high population density and social complexity, are precisely where mobility models are most needed. Yet, our results demonstrate that the EPR model underperforms in these settings. The simplicity of the model struggles to capture the nuanced and constraint-driven behaviors of lower-income individuals in urban areas. This mirrors the WEIRD problem in psychological research, where Western, Educated, Industrialized, Rich, and Democratic populations are overrepresented, leading to skewed findings \cite{muthukrishna2020beyond}. Similarly, the EPR model appears to best describe a particular subset of the population—those with more time and financial resources to explore—while failing to capture the mobility of those constrained by time and money.

These findings have critical implications for the equitable application of mobility models in urban planning. Policies and decisions informed by models that inadequately represent lower-income, urban populations risk exacerbating existing inequalities. Investments and planning of essential services like public transportation infrastructures might overlook the needs of those most reliant on these services if their mobility patterns are not accurately modeled. To address these shortcomings, further research is needed to refine mobility models and ensure they account for the diversity of human behavior, particularly in urban environments.

\section*{Methods} 

\textbf{Data.} Our geo-location data, provided by Cuebiq, consists of anonymized GPS records ("pings") from users who opted in through a GDPR and CCPA-compliant framework. Collected in 2017 under Cuebiq's Data for Good program, the data is used for academic research and humanitarian purposes. The dataset includes pings from 11 core-based statistical areas (CBSAs) between October 2016 and March 2017: New York, Los Angeles, Chicago, Dallas, Philadelphia, Washington, Miami, Boston, San Francisco, Detroit, and Seattle. The original 70.2 billion pings from 14.3 million smartphones were filtered to 67 billion pings from 4.5 million smartphones by considering only devices with over 2,000 pings. Using the Hariharan and Toyama algorithm \cite{hariharan2004project}, we extracted stays from location trajectories and identified venues within a 200-m radius. Stays under 5 minutes or over 1 day were discarded. Home locations were inferred from the most common location (expressed in terms of Census Block Group) between 10:00 p.m. and 6:00 a.m., further refining the data to 976 million stays from 3.6 million individuals. Excluding users with less than 10 nights spent in their inferred home location or without visits to our set of venues, we are left with 1.9 million users. Finally, discarding the lowest quintile in terms of the number of distinct visited places, the final dataset includes visits from 1.5 million users with at least 11 unique visited places. Refer to \cite{moro2021mobility} for full details on privacy protection, visits extraction, and post-stratification techniques.

\textbf{Exploration tendency.} The exploration tendency parameter $\rho_u$ of a user $u$ is given by:

$$
\rho_u = \frac{S_{u} ^ {\gamma + 1}}{  (\gamma +1 ) N_{u} },
$$
where $S_{u}$ is the total number of distinct places and $N_{u}$ is the total number of places visited by the user.

\textbf{Expected inter-event time.} Since at every step the probability of visiting a new place for a user $u$ is given by $P_u (S) = \rho_u S ^ {-\gamma}$, the probability that the next exploration step will occur after $T$ steps is given by:

$$
P(\tau_{u,S} = T) = (1 - P_u (S) )^{T-1} P_u (S),
$$
which is a geometric distribution with expected value $\ev{\tau_{u,S}}$ given by $\ev{\tau_{u,S}}= P_u^{-1} (S) = S ^ {\gamma} / \rho_u$.

\textbf{Sociodemographic indicators.} Each user in our data set is assigned a Census Block Group, as measured
by the 2012–2016 5-year American Community Survey \cite{acs} (ACS), as an inferred home location, using its most common location between 10:00 p.m. and 6:00 a.m. From the same ACS data, we assign the user sociodemographic variables taken or calculated from its home Census Block Group indicators:

\begin{itemize}
    \item Income: median household income
    \item Car, Public transport: fraction of the population who take a given means of transportation (we consider only car and public transport) to commute
    \item Education: fraction of the population with a higher education degree
    \item White, Black, Asian, Hawaiian, Other races: fraction of the population who identifies with a given ethnicity or race.
\end{itemize}

\textbf{Life habit indicators.} Visits in the clean data set are associated with a Foursquare venue and its category. Venue categories are taken from the Foursquare classification \cite{foursquare}. Moreover, categories have been manually grouped into 15 macro groups \cite{moro2021mobility}:  Art/Museum, City/Outdoors, Coffee/Tea, College, Entertainment, Food, Grocery, Health, Service, Shopping, Sports, Transportation, Work. We assign each user a category score for each of the 592 Foursquare categories, based on the fraction of visits spent in places classified in each category.

\textbf{Regression model.} To test the bias of the EPR model towards sociodemographic and life habit indicators, we run a regression model using either $\epsilon_u$ or $\eta_u$ as the dependent variable and the full list of sociodemographic and life habit variables as independent variables. Moreover, we also include $S_u$ and $K_u$, respectively for $\epsilon_u$ and $\eta_u$, as control variables, as they indicate the number of elements in the sum of either $\epsilon_u$ or $\eta_u$, respectively. Given the high number of features (one control, 10 sociodemographic, 592 for life habits), we consider a Least Absolute Shrinkage and Selection Operator (LASSO) regression, whose objective function is:

$$
\min_{\beta} \left( \sum_{i=1}^{n} (y_i - \beta_0 - \sum_{j=1}^{p} \beta_j x_{ij})^2 + \alpha \sum_{j=1}^{p} |\beta_j| \right),
$$
where $y_i$ is the dependent variable, $\beta_i$'s are the coefficients, $x_{ij}$ are the independent variables and $\alpha$ is the regularization parameter. There are numerous reasons to use this type of model in our case. Indeed, through the regularization parameter $\alpha$, it performs automatic feature selection by shrinking the least important coefficients to zero, thus handling multicollinearity and preventing overfitting at the same time. We estimate the parameter $\alpha$ through 10-fold cross-validation with 3 repeated randomizations. All variables, including the dependent ones, have being standardized by substructing the mean and scaling to unit variance. 

\textbf{Exploration burstiness.} Having the sequence of exploration inter-event times $\{\tau_{u,S}\}$ of a user $u$ with $N_u$ total visits, we can compute the burstiness coefficient $B_u$ for finite sequences as defined in \cite{kim2016measuring}:

$$
B_u = \frac{\sqrt{N_u+1}r - \sqrt{N_u-1}}{\left( \sqrt{N_u+1}-2\right)r + \sqrt{N_u-1}  },
$$
where $r = \sigma(\tau_{u,S}) / \mu(\tau_{u,S}) $ is the coefficient of variation ($\sigma$ and $\mu$ are, respectively, the empirical standard deviation and mean of the $\{\tau_{u,S}\}$ sequence. 

\textbf{Preferential return error.} To measure the extent to which a user respects the preferential return principle, we measure individually the empirical deviation from the equation $\Pi_i = \phi_i$, where $\Pi_i$ is the probability of returning to a location $i$ and $\phi_i$ is the fraction of previous visits to that location. To do so, we bin $\phi$ (which is a continuous variable defined in $(0,1]$) in 20 discrete and equally spaced $\phi_b$ intervals ($(0.,0.05]$, $(0.05,0.1]$, ..., $(0.95,1]$), and for every user we build a list of empirical realizations $\Pi_{\phi_b}$ for every bin $\phi_b$, in the following way: at every return step, we recompute $\phi_i$ for each previously visited location $i$ and map it to the respective bin $\phi_b$; then, we add a 1 to the $\Pi_{\phi_b}$ list of the location to which the user actually returns, and 0 to the lists of all the others (if any). After repeating the procedure for all return steps, we compute the empirical probability $\Tilde{\Pi}_{\phi_b}$ as the empirical average of the binary values in $\Pi_{\phi_b}$, for every bin $\phi_b$. $\Tilde{\Pi}_{\phi_b}$ indicates the observed fraction of times the user returned to places previously visited with frequency $\phi_b$. Finally, we can compute how different $\Tilde{\Pi}_{\phi_b}$ is from $\phi_b$, defining the preferential return error $P.R.E.$ as the symmetrical mean absolute percentage error (SMAPE) of $\Tilde{\Pi}_{\phi_b}$ from the expected value $\phi_b$:
 $$
 P.R.E. = \frac{1}{20} \sum_{b=1}^{20} \frac{| \Tilde{\Pi}_{\phi_b} - \phi_b | }{ |\Tilde{\Pi}_{\phi_b}| + |\phi_b|  }.
 $$

\paragraph{Acknoledgements} We would like to thank Cuebiq who kindly provided us with the mobility data set for this research through their Data for Good program. L.N. acknowledges support from the Accelnet-Multinet project. M.K. acknowledges support from the DataRedux Agence Nationale de la Recherche (ANR) project (ANR-19-CE46-0008), the SoBigData++ H2020 project (SoBigData++ H2020-871042), the MOMA WWTF project, and the National Laboratory of Health Security (RRF-2.3.1-21-2022-00006). 

\clearpage

\newpage

\begin{center}
    \Huge Supplementary Information \\

\vspace{5mm}
    
\huge One rule does not fit all: deviations from universality in human mobility modeling

\vspace{5mm}

\LARGE Ludovico Napoli, M\'arton Karsai, Esteban Moro
\end{center}

\newpage

\begin{figure*}[ht!]
  \includegraphics[width=\linewidth]{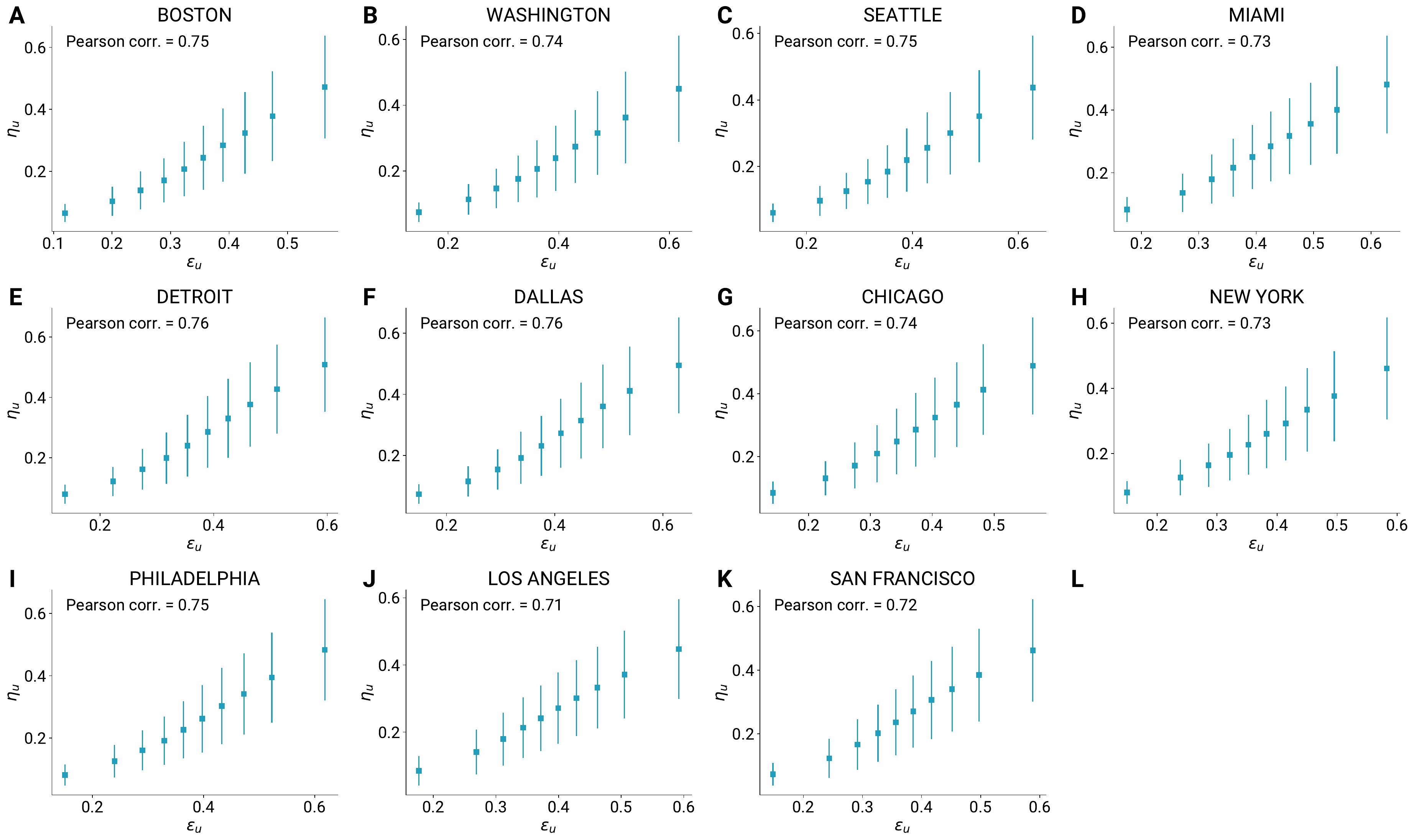}
\caption{The relation between $\epsilon_u$ and $\eta_u$ in all CBSAs.}
\label{figs9}
\end{figure*}

\section*{Deviations}

As mentioned in the main text, the two deviation metrics $\epsilon_u$ and $\eta_u$ are highly correlated, despite measuring aspects of the EPR model apparently uncorrelated. Indeed, $\epsilon_u$ measures how good the EPR model is in predicting in how much time the next exploration step is going to happen. On the other hand, $\eta_u$ measures how "far" is the final visitation frequency distribution from the expected $\ev{f_k} \sim k^{-\gamma}$. The relation between the two variables is shown in Fig.\ref{figs9}, for all CBSAs. The Pearson correlation coefficients go from 0.71 in Los Angeles to 0.76 in Dallas and Detroit, indicating that if a user's exploration dynamics is not well described by the EPR model, then also its visitation frequency is likely not well described.

\begin{figure*}[ht!]
  \includegraphics[width=\linewidth]{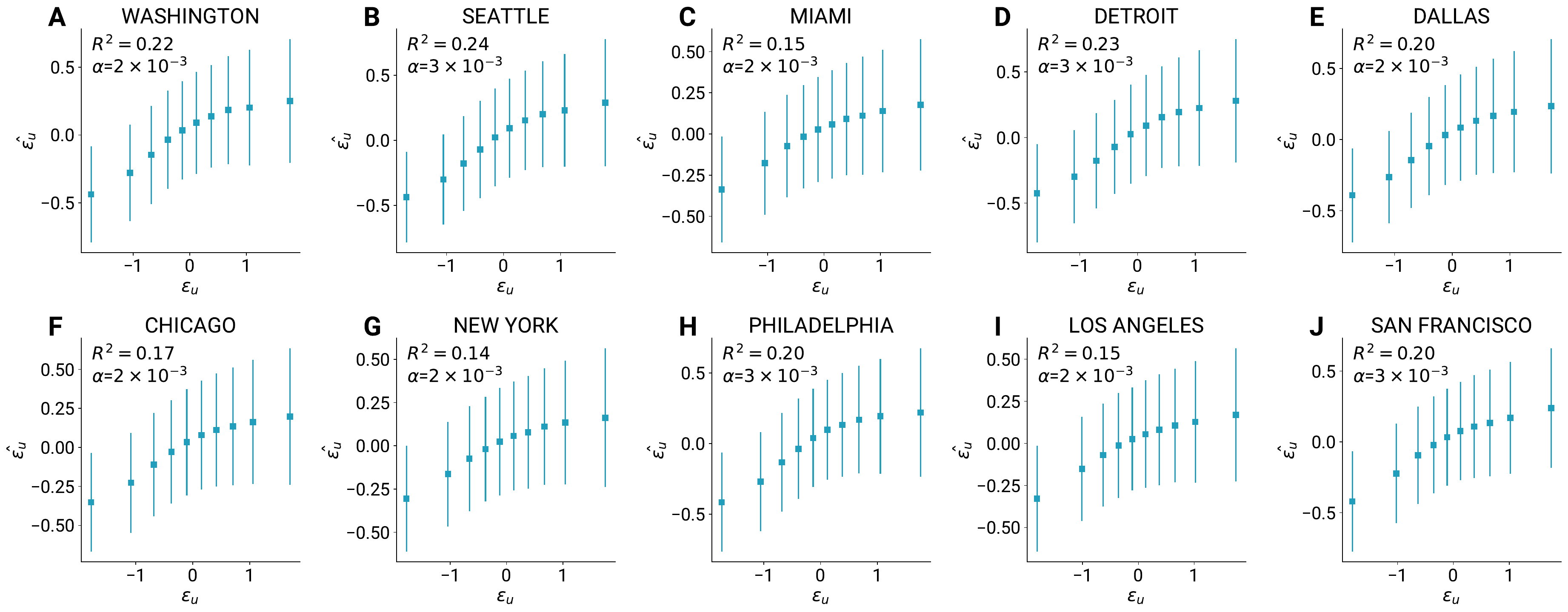}
\caption{Results of the LASSO regression for the other CBSAs (dependent variable: $\epsilon_u$): true $\epsilon_u$ vs predicted $\hat{\epsilon_u}$. $R^2$ is the coefficient of determination while $\alpha$ is the regularization parameter.}
\label{figs5}
\end{figure*}

\begin{figure*}[ht!]
  \includegraphics[width=\linewidth]{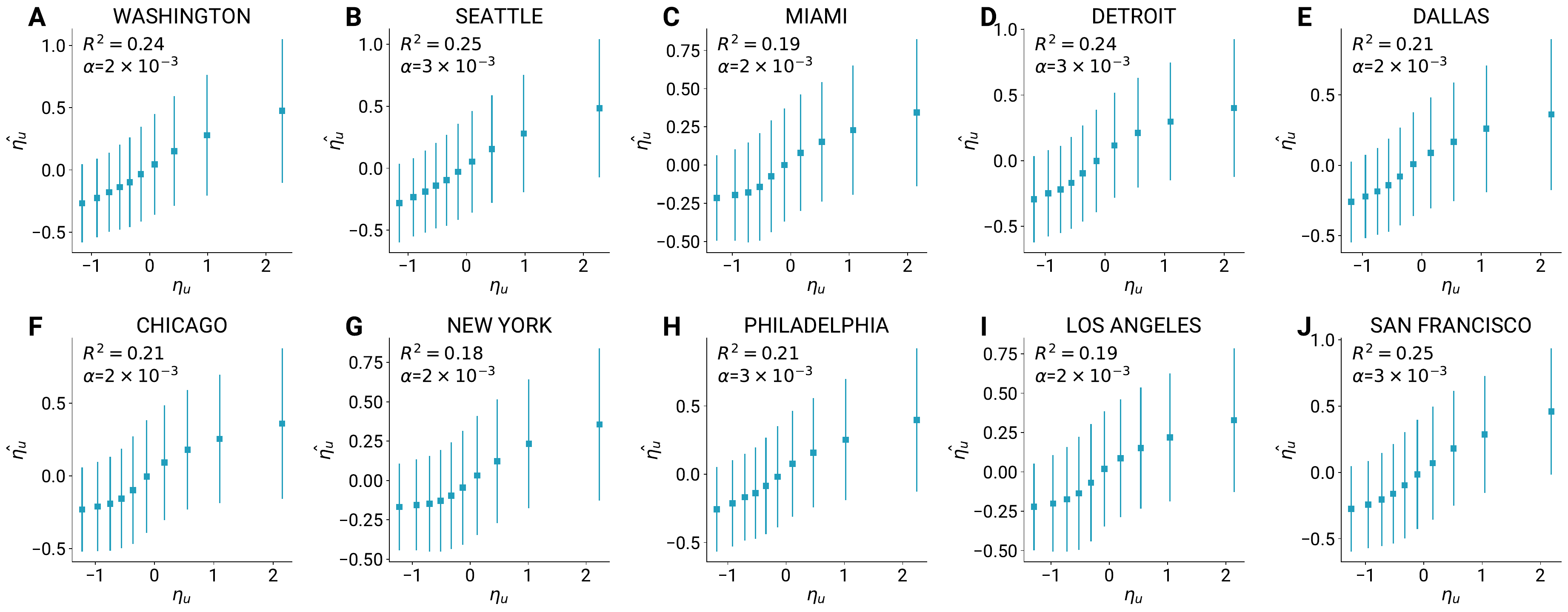}
\caption{Results of the LASSO regression for the other CBSAs (dependent variable: $\eta_u$): true $\eta_u$ vs predicted $\hat{\eta_u}$. $R^2$ is the coefficient of determination while $\alpha$ is the regularization parameter.}
\label{figs6}
\end{figure*}

\begin{figure*}[ht!]
  \includegraphics[width=\linewidth]{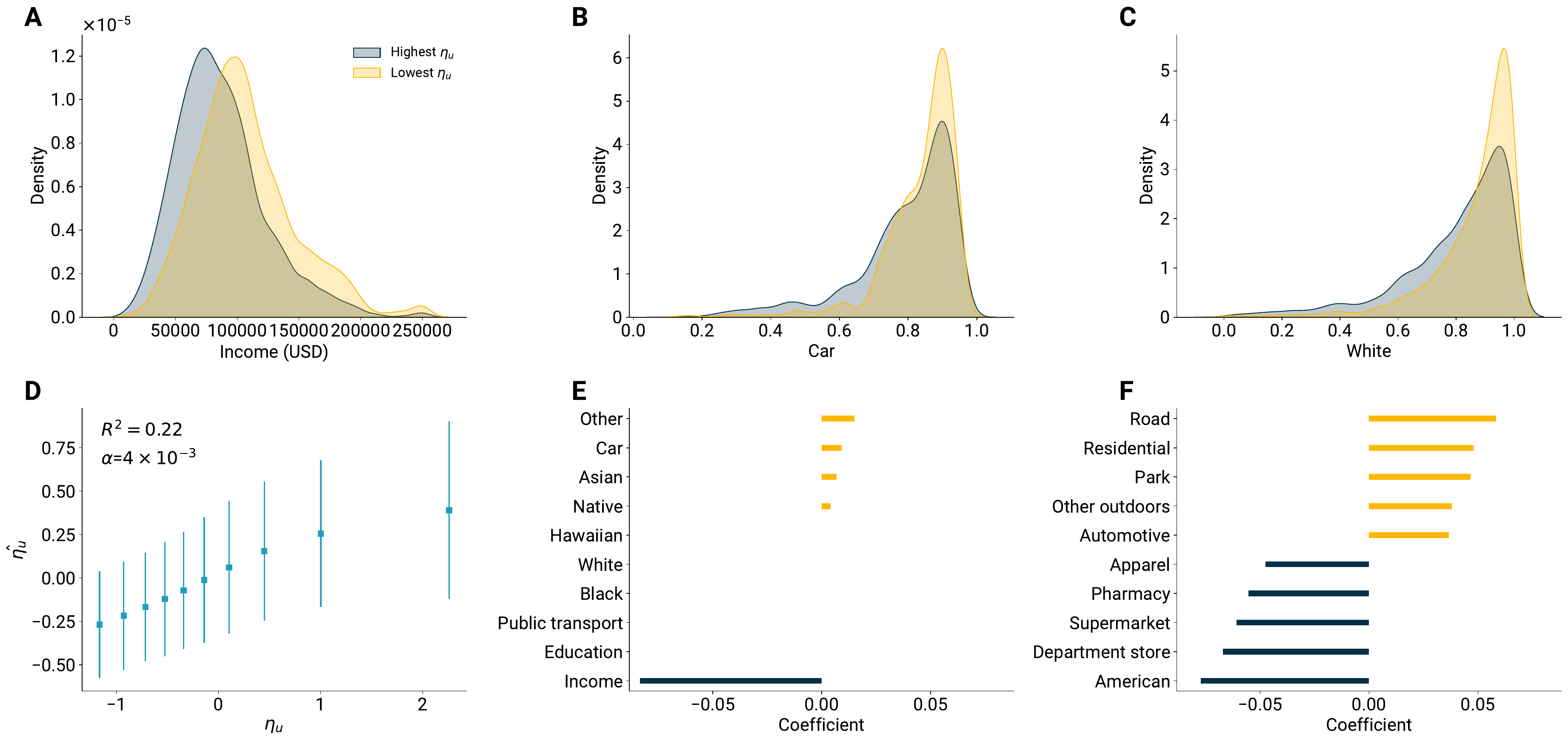}
\caption{A) The income distribution of users in the highest (blue) and lowest (yellow) 10\% quantile of $\eta_u$. B) Same as A) for the use of car. C) Same as A) for the probability of being white. D) Results of the LASSO regression: true $\eta_u$ vs predicted $\hat{\eta_u}$. $R^2$ is the coefficient of determination while $\alpha$ is the regularization parameter. E) Coefficients of the socioeconomic features. F) Coefficients of the life habit features (only highest and lowest 5 shown).}
\label{figs1}
\end{figure*}

\section*{Regression models}

As mentioned in the main text, both $\epsilon_u$ and $\eta_u$ can be partially predicted from sociodemographic and life habits features, indicating that the EPR model is biased towards certain groups of people. We perform a LASSO regression for both deviation variables, estimating the regularization parameter $\alpha$ through cross-validation (see the Methods section in the main text for details), to quantify this bias in every CBSA. The regression results for $\epsilon_u$ are shown in Fig. \ref{figs5}, where we show the relation between the actual $\epsilon_u$ and the predicted $\hat{\epsilon_u}$. The coefficient of variations goes from $R^2=0.14$ for New York to $R^2=0.24$ for Seattle. The results for $\eta_u$, on the other hand, are shown in Fig. \ref{figs6}, where we show the relation between the actual $\eta_u$ and the predicted $\hat{\eta_u}$. The coefficient of variations in this case goes from $R^2=0.18$ for New York to $R^2=0.25$ for Seattle and San Francisco. The results show the bias towards sociodemographic and life habit variables is robust and consistent across all CBSAs.

\section*{Results for $\eta_u$ in Boston}

The analogous results of Fig.2 in the main text for $\eta_u$ are shown here in Fig.\ref{figs1}. As we can see in the figure, results are consistent with $\epsilon_u$, with the only significant exception being the use of car, which is negatively associated with $\epsilon_u$ and positively with $\eta_u$.

\begin{figure*}[ht!]
  \includegraphics[width=\linewidth]{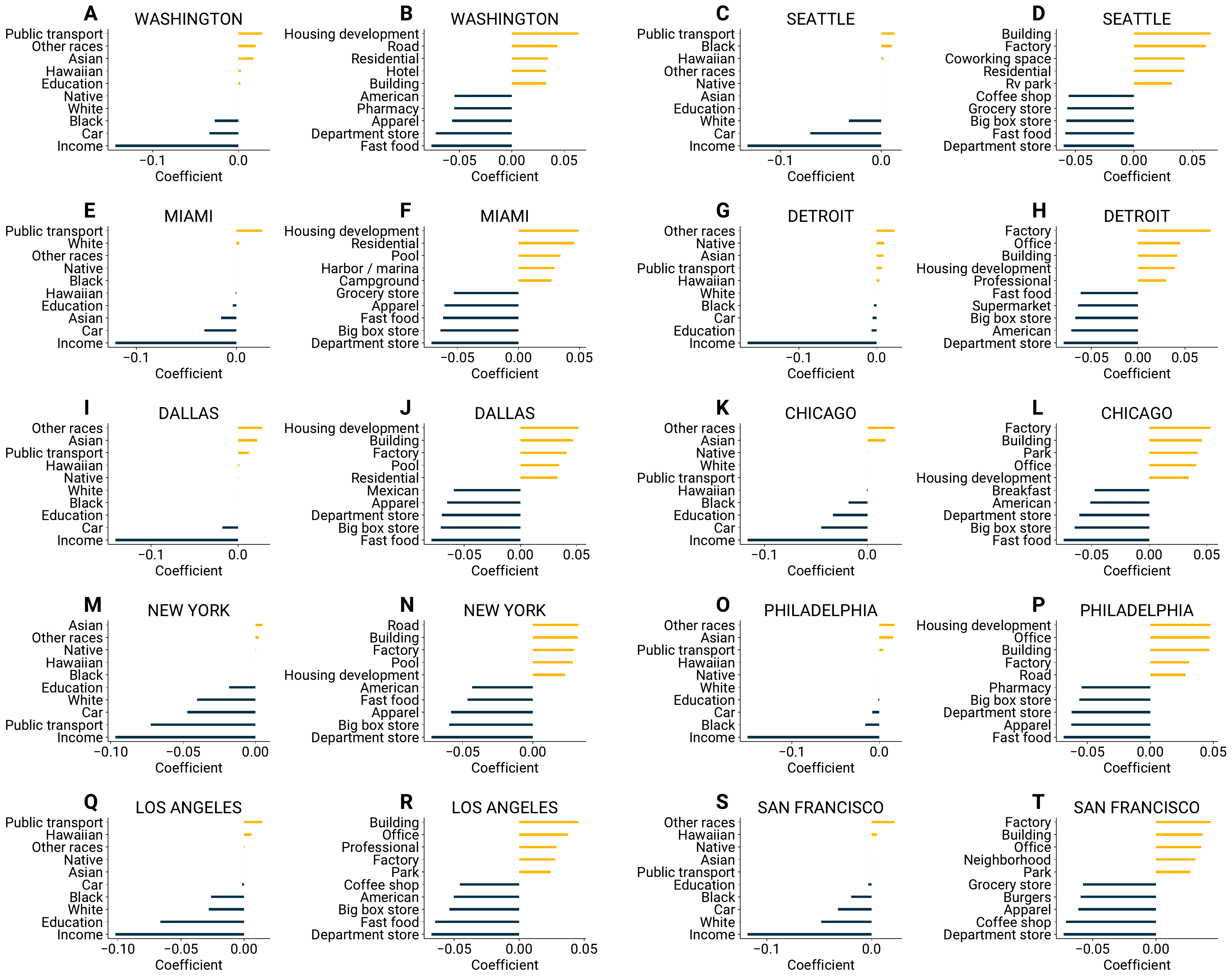}
\caption{First and third column: Coefficients of the socioeconomic features for the other CBSAs (dependent variable: $\epsilon_u$). Second and fourth column: Coefficients of the life habit features (only highest and lowest 5 shown) for the other CBSAs (dependent variable: $\epsilon_u$.)}
\label{figs7}
\end{figure*}

\begin{figure*}[ht!]
  \includegraphics[width=\linewidth]{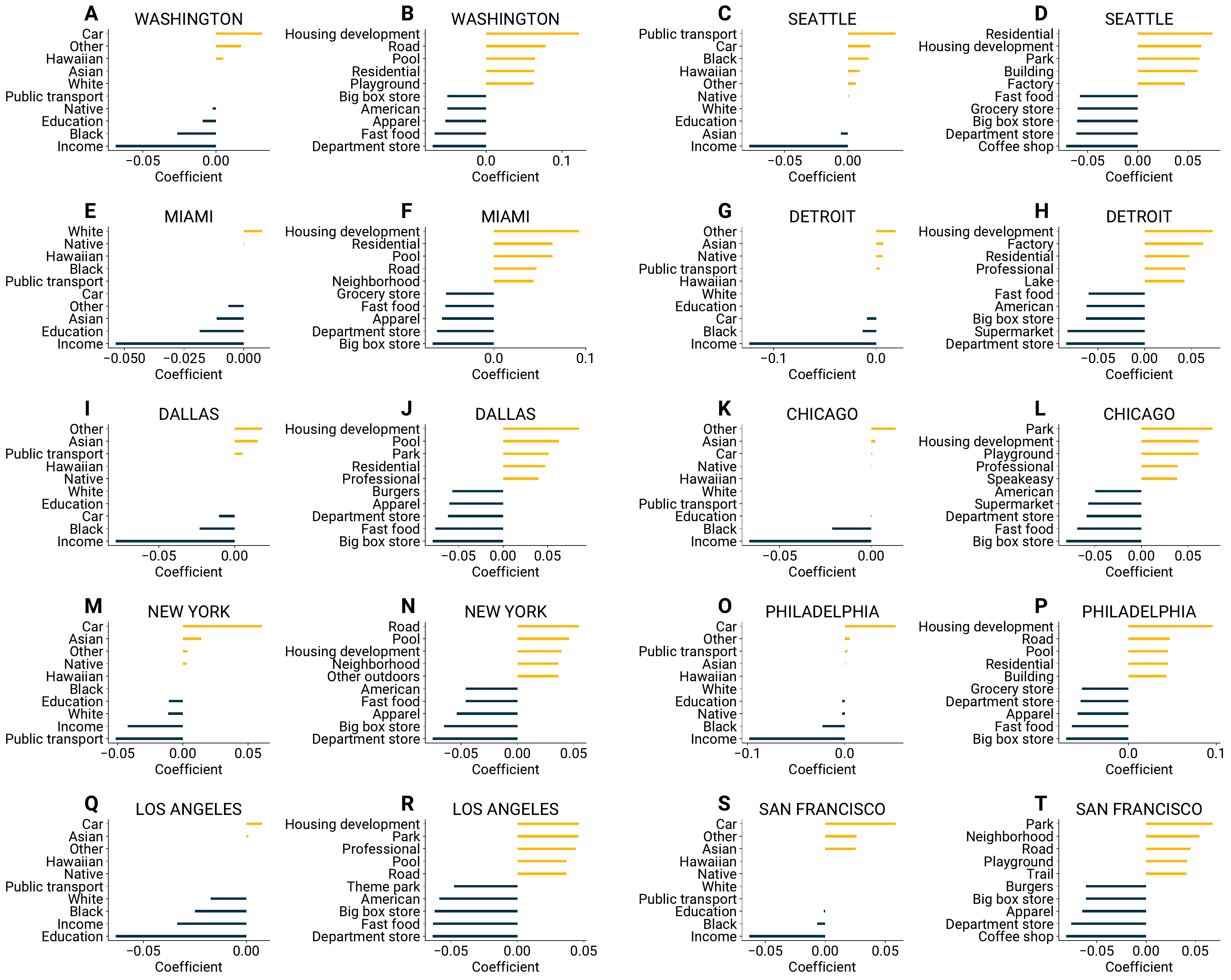}
\caption{First and third column: Coefficients of the socioeconomic features for the other CBSAs (dependent variable: $\eta_u$). Second and fourth column: Coefficients of the life habit features (only highest and lowest 5 shown) for the other CBSAs (dependent variable: $\eta_u$.)}
\label{figs8}
\end{figure*}

\section*{Feature importance}

In the main text, we measure the bias of the EPR model towards sociodemographic and life habit characteristics in Boston through the coefficients of the LASSO regressions. Here we analyze the same results for all the other CBSAs.

In Fig.\ref{figs7} we show the coefficients of the regressions on $\epsilon_u$. As mentioned in the main text, the results are consistent across CBSAs. First, regarding the sociodemographic features (first and third columns in the figure), income is always the most important predictor, with a negative coefficient: the lower the income, the higher $\epsilon_u$. Moreover, also the use of cars has a consistent negative impact on $\epsilon_u$, except for Los Angeles, Philadelphia, and Detroit, where the coefficient is almost zero. Complementary to the use of cars, the use of public transport is either positively associated with $\epsilon_u$ or irrelevant. Education is not always significant, but when it is the coefficient is negative. Finally, the role of race is mostly irrelevant, despite some exceptions: a negative coefficient is observed for the probability of being white in New York, Los Angeles, Seattle, and San Francisco, while it is observed for the probability of being black in Washington, Los Angeles, San Francisco, and Chicago. On the other hand, the probability of being Asian is often positively associated with $\epsilon_u$, though with a small coefficient, e.g. in Washington, Dallas, Chicago, and Philadelphia. 

Regarding the life habit features (second and fourth columns in the figure), we also find some similar results as in the main text for the other CBSAs. Indeed, among the categories of places with the highest positive coefficients, we find mostly routine places like roads, factories, buildings, and offices. On the other hand, among places that are negatively associated with $\epsilon_u$, we find also many restaurants, like fast food and American, and shops, like supermarkets, grocery stores, department stores, and apparel shops.

Regarding the coefficients of the regressions on $\eta_u$, shown in Fig.\ref{figs8}, the only notable difference with the ones for $\epsilon_u$ is the coefficient of the use of cars, which is positive and significant in most places, like we observed for Boston in the main text.

\begin{figure*}[ht!]
  \includegraphics[width=\linewidth]{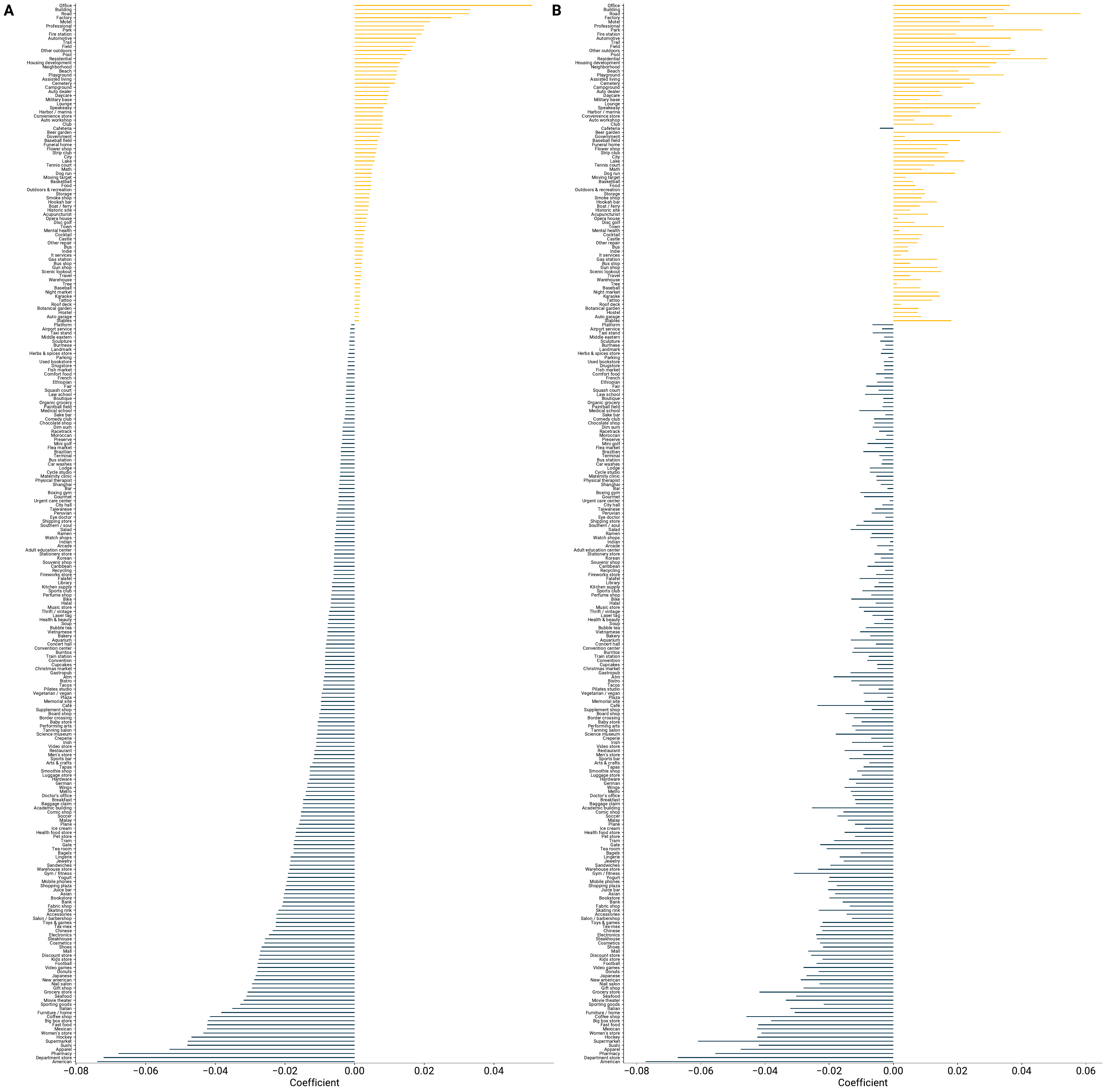}
\caption{A) Coefficients of all the life habit features (dependent variable: $\epsilon_u$), with a coefficient greater than $10^{-3}$ in both regressions. A) Coefficients of all the life habit features (dependent variable: $\eta_u$), with a coefficient greater than $10^{-3}$ in both regressions.}
\label{figs4}
\end{figure*}

\section*{Life habits categories}

In Fig.\ref{figs4}, we show the full list of coefficients of the regression on $\epsilon_u$ (panel A) and $\eta_u$ (panel B) for Boston, that have an absolute value higher than $10^{-3}$ for both regressions. In both panels, coefficients are sorted based on the values of the regression on $epsilon_u$. Notably, the sign of coefficients is always the same for the two regressions, except only for Cafeteria, which has a positive relation with $\epsilon_u$ and a negative one with $\eta_u$. Moreover, in addition to the sign, coefficients also share similar absolute values. In other words, categories that are important for predicting $\epsilon_u$ are likely to be important for predicting $\eta_u$, with the same direction.

As mentioned in the main text and in the previous section, among the categories with the most significant coefficients we find also restaurants, amusement places, e.g. hockey arenas and movie theatres, and shops.

\begin{figure*}[ht!]
  \includegraphics[width=\linewidth]{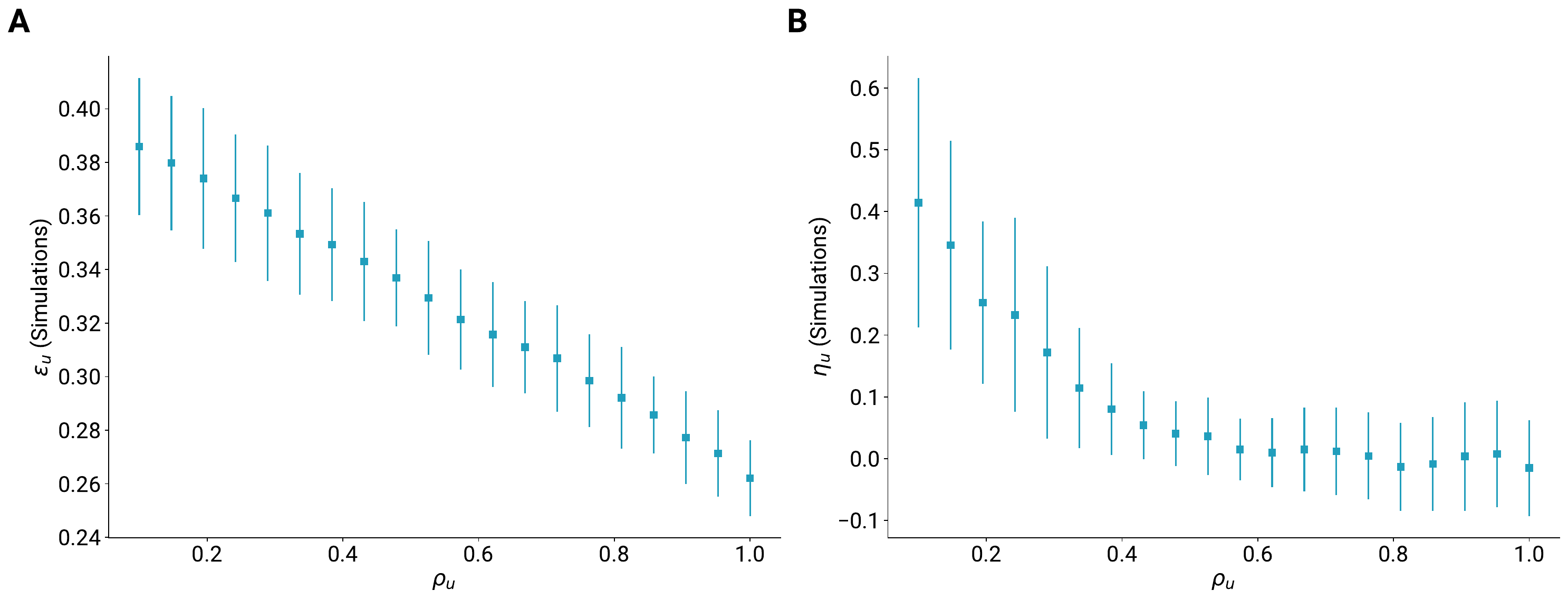}
\caption{A) Relation between $\rho_u$ and $\epsilon_u$ in simulated experiments. B) Relation between $\rho_u$ and $\eta_u$ in simulated experiments.}
\label{figs10}
\end{figure*}

\section*{Stochastic deviations}

The stochastic deviations that we generate from simulations in the orange distributions in panels C-F of Fig.1 in the main text are generated using the individual parameters taken from the data, to make the results directly comparable with the empirical deviations, shown in the light blue distributions in the same figure. To generate a stochastic deviation $\epsilon_u$ for a user $u$, we run a simulation of the EPR model for a fictitious user with the same visitation tendency $\rho_u$ and the same number of distinct places $S_u$ of the real user. On the other hand, to generate a stochastic deviation $\eta_u$, we run another simulation, using the same $\rho_u$ and number of steps $N_u$, then we consider only the visitation frequency of the first $K_u$ distinct places (where also in this case $\rho_u$, $N_u$ and $K_u$ are taken from the data. The details on the simulations can be found in the following paragraphs, with the only difference that instead of having a distribution of parameters taken from the data of individual users, we fix all of them and tune only $\rho_u$, because we are specifically interested in measuring the role of this parameter and to do this we need to control for the others.

As mentioned in the main text, the stochastic errors are related to the exploration tendency, encoded in the parameter $\rho_u$. However, the stochastic deviations that we generate in the orange distributions in panels C-F of Fig.1 in the main text, as we mentioned above, do not only depend on $\rho_u$ but also on other individual parameters like the number of steps $N_u$ and the number of distinct places $S_u$. To demonstrate the dependence on $\rho_u$, then, we run other stochastic simulations of the EPR models where we control for the other parameters, such that the final results can be directly associated with $\rho_u$. We run such controlled experiments for 20 different values of $\rho_u$, keeping all the other parameters fixed.

For $\epsilon_u$, we consider simulations for $S_u=100$ distinct places. As explained in the Methods section in the main text, in the EPR model the inter-event time is drawn from the distribution $
P(\tau_{u,S} = T) = (1 - P_u (S) )^{T-1} P_u (S)
$, where $P_u(S) = \rho_u S^{-\gamma} $ is the user's probability of exploration. In every simulation, we compute the simulation's $\epsilon_u$ as:

$$
\epsilon_u = \frac{1}{S_u} \sum_{S=1}^{S_u} \frac{|\tau_{u,S} - \ev{\tau_{u,S}} |}{|\tau_{u,S}| + |\ev{\tau_{u,S}}|}
$$

where $\tau_{u,S}$ is the random value drawn from $P(\tau_{u,S} = T)$ and $\ev{\tau_{u,S}} = 1 / P_u(S)$ is the expected value. We run 100 simulations for each value of $\rho_u$ and compute $\epsilon_u$ for every simulation. In other words, we consider an ensemble of 100 identical users generated by the EPR model. In panel A of Fig.\ref{figs10} we show the mean and standard deviation of the distribution of $\epsilon_u$ for every value of $\rho_u$. As can be seen in the figure, the values that we get in the simulations decrease linearly with $\rho_u$. The result of these simulations indicates that the stochastic part of $\epsilon_u$ that we measure from data depends on $\rho_u$ and decreases with it.

\begin{figure*}[ht!]
  \includegraphics[width=\linewidth]{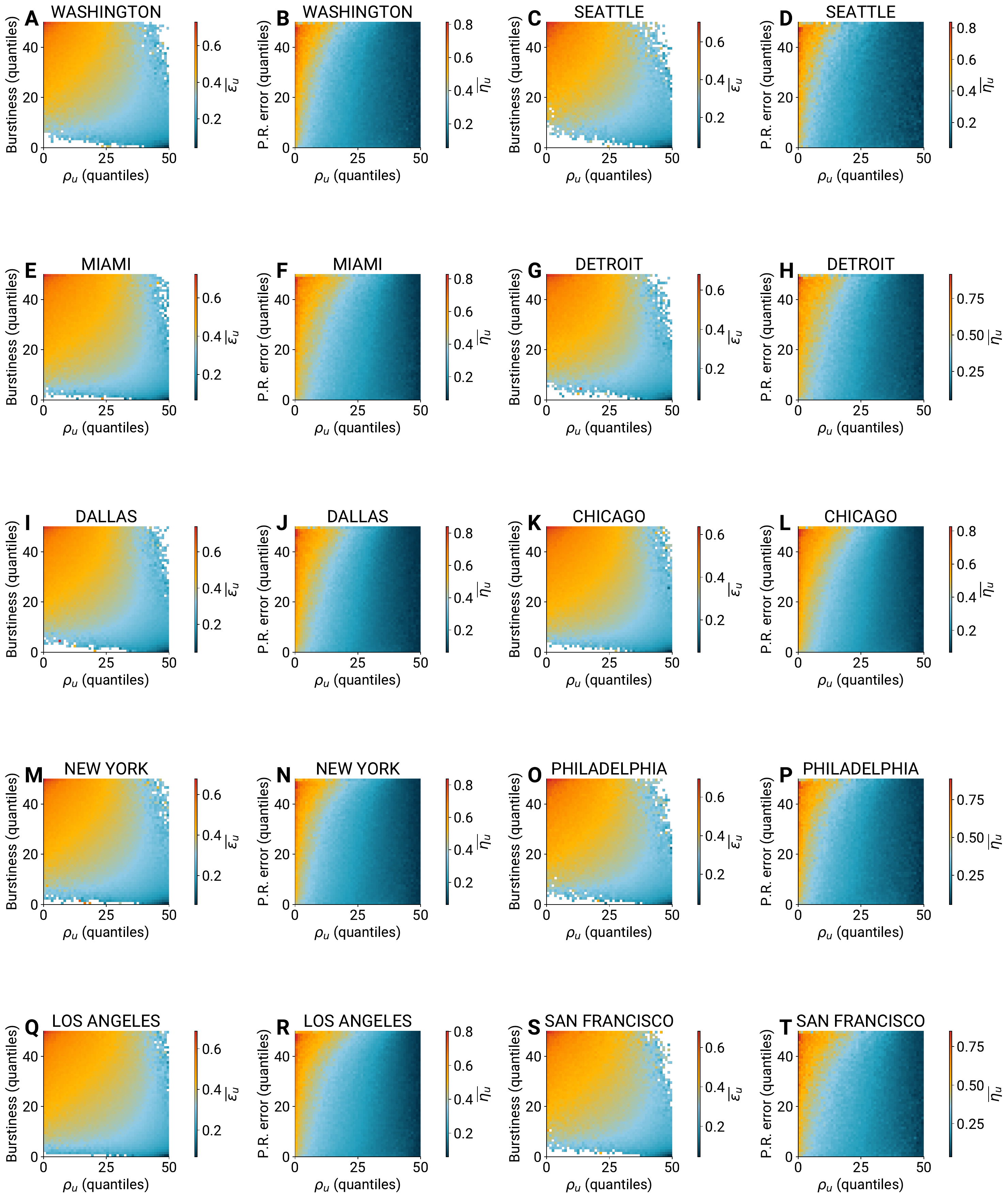}
\caption{First and third column: Average values of $\epsilon_u$ for users grouped in quantiles of $\rho_u$ and burstiness, for the other CBSAs. Second and fourth column: Average values of $\eta_u$ for users grouped in quantiles of $\rho_u$ and P.R. error, for the other CBSAs.}
\label{figs3}
\end{figure*}

For $\eta_u$, we run simulations of 100 steps, where at each step the fictitious user explores or returns with a probability $P_u(S) = \rho_u S^{-\gamma} $. At the end of the simulation, we compute the visitation frequency distribution and compute the $\eta_u$ as the KL-divergence with $f_{u,k} \sim k^{-\gamma-1}$:

$$
\eta_u = \sum_{k=1}^{K_u} f_{u,k} \log \frac{f_{u,k}}{\ev{f_{u,k}}}
$$

where $K_u$ is the rank of the least visited among the locations visited more than once (as explained in the main text, we don't consider locations visited only once to take out tail effects). Also in this case, we run 100 simulations for each value of $\rho_u$ and compute $\eta_u$ for every simulation. In panel B of Fig.\ref{figs10} we show the mean and standard deviation of the distribution of $\eta_u$ for every value of $\rho_u$. Similarly to what we observe with $\epsilon_u$, also $\eta_u$ from simulations decreases with $\rho_u$, although with a sharper decrease for low values of $\rho_u$ and a very smooth decrease for higher values. Also in this case, these simulations indicate that the stochastic part of $\eta_u$ that we measure from data depends on $\rho_u$ and decreases with it.

\section*{Microscopic mechanisms}

In Fig.\ref{figs3} we show the analogous results of Fig.3 A and B in the main text for all the other CBSAs. As we can see from the first and third columns of the figure, the positive association between exploration burstiness and $\epsilon_u$, after controlling for the stochasticity through $\rho_u$, is visible and consistent across all CBSAs. On the other hand, from the second and fourth columns of the figure, we can see that also the role of the P.R. error in determining $\eta_u$ is visible and consistent across all CBSAs, except for the users with an extremely high visitation tendency (similarly to what we have seen for Boston in the main text).

\section*{Characterization of assumptions’ violations}

\begin{figure*}[ht!]
\centering
\includegraphics[width=0.7\linewidth]{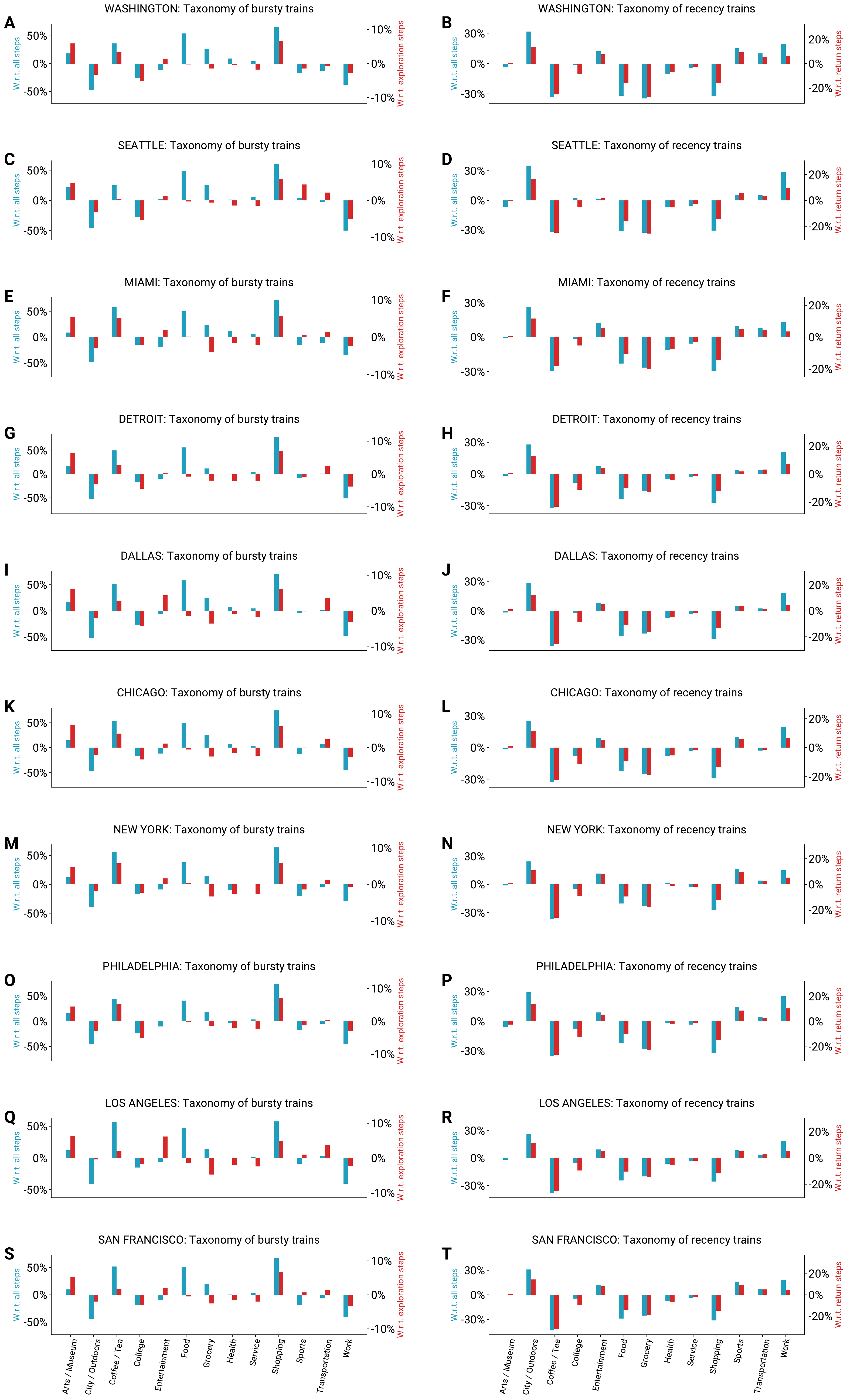}
\caption{Left column: Characterization of bursty trains in all CBSAs, i.e., sequences of consecutive exploration steps, in terms of relative visits to defined categories, compared to all visits (blue bars, left y-axis) and to exploration steps only (red bars, right y-axis). Right column: Characterization of recency trains in all CBSAs, i.e., sequences of consecutive visits to the same place, in terms of relative visits to defined categories, compared to all visits (blue bars, left y-axis) and to return steps only (red bars, right y-axis).}
\label{figs11}
\end{figure*}

In Fig.\ref{figs11} we show the analogous results of Fig.3 C and D in the main text for all the other CBSAs. As can be seen in the figure, bursty trains are indeed characterized by the same specific categories in all CBSAs. Indeed, museums and art galleries, coffee shops, and shopping locations are visited significantly and consistently in all CBSAs more often during bursty exploration trains. Conversely and equally consistently in all CBSAs, workplaces, and venues in the \textit{City / Outdoor} category, such as parks, neighborhoods, and residential places, are rarely explored during bursty trains.

The same mirrored pattern as Boston in recency trains is observed in other CBSAs. In fact, when people repeatedly return to the same place, they tend to do so at routine and habitual locations, such as residential areas, workplaces, as well as transportation hubs and sports venues, and this is consistently true across CBSAs. In contrast and equally consistently, amusement places like coffee shops, restaurants, and shopping malls are not typically revisited continuously.

\section*{Spatial distribution}

\begin{figure*}[ht!]
  \includegraphics[width=\linewidth]{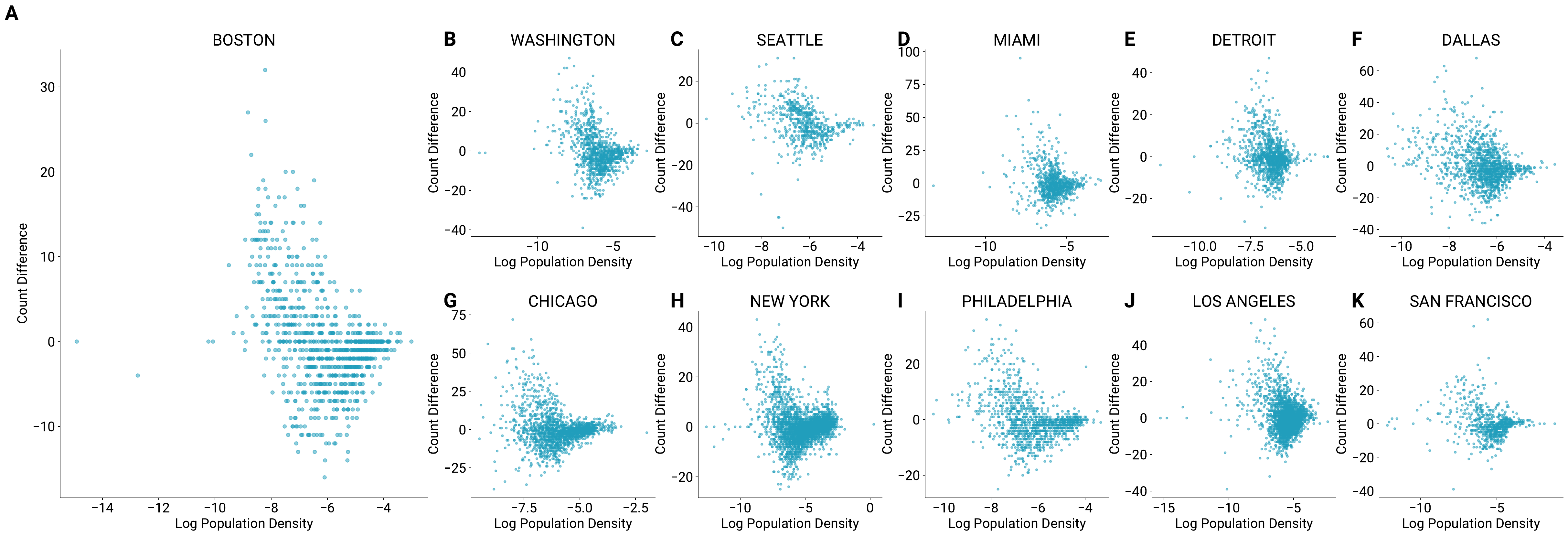}
\caption{Relation between the logarithm of tracts' population density and the difference between the number of users in the highest and lowest 10\% of $\epsilon_u + \eta_u$, for the other CBSAs.}
\label{figs2}
\end{figure*}

As mentioned in the text, we can see a clear urban-rural pattern in all CBSAs, visible in Fig.4 in the main text. Indeed, users who are best represented by the EPR model, i.e. who are in the lowest 10\% of the distribution of the combined deviation variable $\epsilon_u + \eta_u$, are mostly located far from the city center, in bigger and less urban census tracts. On the other hand, the most non-well-represented users, i.e. who are in the highest 10\% of the distribution of the combined deviation variable $\epsilon_u + \eta_u$, are mostly located closer to the city center, in smaller and more urban census tracts, though in the very central tracts, the two quantiles are mostly balanced. These visual results are confirmed by the more robust analysis shown in Fig.\ref{figs2}, where we explicitly show the relation between a tract's population density (a proxy to its urbanization level) and the variable shown in the maps of Fig.4 in the main text, i.e. the count difference between the two extreme quantiles. The relation is negative in all CBSAs, with Pearson correlation coefficients that go from -0.23 in New York and San Francisco to -0.37 in Dallas and Philadelphia and to -0.41 in Boston, as shown in Table 1 in the main text.

\clearpage




\end{document}